\documentstyle[12pt]{article}
\begin{document}
\baselineskip .33in
\pagestyle{empty}
\begin{center}

{\Large {\bf Extended electronic states in  disordered  1-d  lattices:
                an example }} \\
\vskip 1cm
S. Sil, S. N. Karmakar and R. K. Moitra  \\
\vskip .8cm
{\em Saha Institute of Nuclear Physics  \\
1/AF, Bidhannagar, Calcutta 700 064, India} \\
\vskip 1.5cm
{\bf Abstract }
\end{center}

          We discuss a very simple model of a 1-d  disordered  lattice,  in
which {\em all}
the electronic eigenstates are  extended.  The  nature  of  these
states is examined from several  viewpoints,  and  it  is  found  that  the
eigenfunctions are not Bloch functions although they extend throughout  the
chain. Some typical wavefunctions are plotted. This problem  originated  in
our earlier  study of extended states  in  the  quasiperiodic  copper-mean
lattice [ Sil, Karmakar, Moitra and Chakrabarti, Phys. Rev. B (1993)  ].  In
the
present investigation  extended states are found to arise from a  different
kind of correlation than that of the well-known dimer-type.

\newpage
\pagestyle{plain}
\setcounter{page}{1}
\sloppy
Following the original work of Anderson  \cite{ref1},  there  is  a  vast
literature on the nature of electronic eigenfunctions in disordered systems
\cite{ref2}. Over the past decade,  the  scaling  approach  has  been
successfully
utilised to elucidate the problem of  the  dimensional  dependence  of  the
localisation and related problems in such systems [3-5]. It is now part  of
the folklore that in 1-d systems all the electronic  states  are  localised
for any type of disorder \cite{ref5,ref6}, be it  chemical  or  topological.
In  this
letter we provide a very simple example of a random system in which all the
states   are   extended,   contrary    to    popular    belief,    although
none of  the eigenfunctions is of Bloch type.

Instances,  of  course,  are  available  in  the  literature   of
disordered or aperiodic systems in which extended states are supported at a
few isolated  energies  [7-13].  An  well-known  example  is  the  randomly
distributed dimers on a host lattice \cite{ref9}, in which the contribution
to  the
total transmission matrix due to a dimer can be shown to be identity for an
energy equal to the site-energy of a constituent atom of the dimer
\cite{ref14}. At
this energy, therefore, the system supports an extended state provided this
energy lies  within  the  host  band.  Another  example  is  the  aperiodic
copper-mean lattice \cite{ref13}, recently studied by us,
which has been  shown  to
possess mini-bands of extended  states.  In  this  letter  we  provide  yet
another instance of a sytem, a random one this time, which surprisingly has
{\em all}
its  states  extended,  but  none  of  which  is  of  Bloch-type.  The
hamiltonian of this system is given by
\begin{equation}
              H = \sum_{i} \varepsilon_{i} |i\rangle \langle i| +
\sum_{<ij>} t_{ij} |i\rangle \langle j|
\end{equation}
where the hopping matrix element assumes the  values  $t$  and
$-t$, the concentrations of which are $x$ and $1-x$ respectively,  all  the
site-energies having the same value $\varepsilon$. Interestingly,
this random  hamiltonian
can be trivially mapped onto an ordered hamiltonian with $t$  as  the  value
of
hopping matrix  element  by   a   similarity   transformation   $S$,   which
has
elements $+1$ and $-1$ placed  appropriately   along   the   diagonal,   all
other
elements being equal to zero. This immediately implies that the  eigenvalue
spectrum of this random hamiltonian is  identical  with  the  corresponding
periodic chain. So the density of states of these  two  systems  are  the
same, and therefore by the well-known formula due  to  Thouless
\cite{ref15},  the
localisation length is infinity. Thus all states in this random system  are
extended. This argument is independent of  the  concentration  of  the  two
kinds of bond on the lattice, and of any possible short  range or long
range correlation among the bonds.

A very simple argument based on the  real  space  renormalisation
group method \cite{ref13}
immediately shows why the states  are  extended  in  this
sytem. Following the usual method due to Southern et al. \cite{ref16},
if  alternate
sites on a linear chain having a  distribution  of  bonds
$+t$  and  $-t$  are
eliminated, the resulting chain has renormalised site  energies  and  bonds
given by
\begin{equation}
\varepsilon ' = \varepsilon + \frac{2t^2}{E-\varepsilon }~~~~~\mbox{\rm and}
{}~~~~~t' = \pm ~ \frac{t^2}{E-\varepsilon}
\end{equation}
Since the renormalised site energies involve the square of  $+t$  or
$-t$, we have a renormalised chain in which all site energies are equal  and
there are again two types of hopping integrals having  the  same  magnitude
but differing in sign. Also, the site energy  and  the  magnitudes  of  the
hopping integrals are the same as the  corresponding  renormalised  ordered
chain. Thus the evolution of the parameters exactly parallels that  of  the
ordered chain; upon iteration  it  is  found  that  both  the  renormalised
hopping integrals $t$ and $-t$ keep
on oscillating without decaying   for  real
energies, just  as  in  the  ordered  case.  This  indicates  the  extended
character of the eigenstates. The identity of spectra of  the  two  systems
also follows from this analysis.

Yet another way to look at  the  problem  would  be  through  the
transfer matrix approach \cite{ref13,ref14}.
In a disordered chain of $+t$ and $-t$ bonds,
we  may  have  eight  possible  configurations  of  triplets   of   bonds,
corresponding to assigning either the value $+t$ or $-t$
to every bond. It  can
be  easily  checked  that  the  composite  transfer  matrices   for   these
configurations turn out to be either $\pm I$ (Identity  matrix) or
$\pm \sigma_z$ (z-component of Pauli matrix)
at an energy value $E = \varepsilon$. Clearly then,
for this value  of  energy  the
total transmission matrix is a product of a string of  matrices  consisting
of $I$ and $\sigma_z$,  and  therefore  the  total  transmission
coefficient is unity.  If we now renormalise the chain, then applying these
considerations again to this chain, it is easy to see that at enegies
$E=\varepsilon \pm \sqrt{2} t$, which are the solutions of the equation
$E  = \varepsilon^{(1)}(E)$, the  total  transmission
coefficient is  unity.
Repeated renormalisation of
the chain yields as roots of the equation $E = \varepsilon^{(n)}(E)$,
the energy eigenvalues
\begin{equation}
E^{(n)} =\varepsilon \pm t\sqrt{2\pm \sqrt{2\pm \sqrt{2\pm \cdots
\mbox{\rm n times}}}}
\end{equation}
which are the same as those for an ordered chain with hopping matrix
elements $t$ \cite{ref13}.
These eigenvalues span, in the limit $n \rightarrow \infty$,
the  full  spectrum  of  energies
ranging from $-2t$ to $2t$.
It  follows  therefore  that  for  all
energies within the band  the  transmission  coefficient  is  unity,  which
confirms that {\em all}
the eigenstates in the random chain are extended.

It is interesting to consider for the sake  of  illustration  the
wavefunctions for a few selected cases. The hamiltonian parameters, in
all these cases, are taken as $\varepsilon =0$ and $|t| = 1$, and the
wavefunctions are plotted at an energy value $E=1.99$.
Consider a chain in which only  one
bond has the value $-t$, all other bonds being  $t$.
The wavefunction  in  this
case is of the ``domain wall'' type, and is shown  in  fig.1(b),  where  we
have also given  the  wavefunction  for  the  ordered  chain  alongside  in
fig.1(a) for the sake of comparison. In  this  and  subsequent  figures  we
display the amplitudes only on the first 150 sites of a chain of length  of
the order of 5,00,000 bonds. As the overall
characterstics of the wavefunctions are found to
remain the same throughout the  chain,  the
amplitudes on the rest of the sites are not shown. In
fig.1(c) we display the wavefunction for a chain  in  which  bond  strengths
$t$
and  $-t$  appear  randomly  with   equal   probability.    A   particularly
interesting
case is the bond model of the Fibonacci lattice for which  all  the  states
are known to be critical \cite{ref17}.
However, if we assign the special  values  $t$
and $-t$ to the ``long'' ($L$) and ``short'' ($S$) bonds in this model,
all the states become
extended. In fig.1(d) we have plotted  a  typical  wavefunction  for  this
chain. The interesting feature common to  figures  1(b)-1(d)  is  that  the
amplitudes oscillate without decaying from one end  of  the  chain  to  the
other, although there is no periodicity. Thus these are not Bloch states.

The reason for the extended nature of the eigenstates  in
the Fibonacci case becomes at once apparent  if one looks at the expression
for the invariant associated with the bond model of this chain \cite{ref17}
\begin{equation}
{\cal I} =\frac{1}{4} \left ( \frac{t_S}{t_L} - \frac{t_L}{t_S} \right )^2 .
\end{equation}
This, in addition to becoming zero for $t_L = t_S = t$,
corresponding  to  the
ordered limit, {\em also} is seen to become zero for $t_L = -t_S = t$,
a  fact  which
surprisingly has gone unnoticed so far in  the  literature.

The problem discussed in this letter arose while considering  the
nature  of the extended states in a copper-mean  aperiodic  chain,  studied
recently by us using the real space renormalisation group  method
\cite{ref13}.  A
general method has  been  proposed  for  determining  the  eigenvalues  and
eigenfunctions of the  extended  states  in  such  systems,  where  clearly
Bloch's theorem is of no help.  The hamiltonian parameters  were  found  to
display a characteristic pattern at  a  certain  stage  of  renormalisation
depending on the energy eigenvalue for an extended state, namely  that  all
the site-energies became equal and the  long  bonds  and  the  short  bonds
assumed  equal and opposite values. This  problem  thus  becomes  identical
with the one discussed in this paper.

An additional insight that we gain from this work  is  that
extended states are seen to be possible in random systems  not  necessarily
restricted to possessing dimer-type correlations  \cite{ref9,ref13};
as  we  have  shown
above, it is sufficient to have  a  string  of  $I$  and  $\sigma_z$
operators in the total transmission matrix  at  certain  energy  values  in
order make the total transmission coefficient  equal  to  unity.  This  can
arise in a more general type of situation than the  one  in  which    dimer
type correlations alone is present.

\newpage
\noindent
{\Large {\bf Figure Captions }}
\begin{itemize}
\item[Figure 1] Amplitudes of the wavefunction on the first 150 sites of
a chain of 5,00,000 bonds for (a) an ordered chain with $t = 1$, (b)
a chain with $t=-1$ for the 62nd bond and with $t=1$ elsewhere, (c) a chain
on  which  bonds  assume  values  $t=1$  and  $t=-1$  randomly  with   equal
probability,
and (d) a fibonacci chain with $t_L=1$ and $t_S=-1$. The energy eigenvalue
chosen is $E=1.99$ and the site-energy is $\varepsilon =0$, both measured in
units of $|t|$. The amplitudes on the zeroth and the first sites have been
taken to be 0 and 1 respectively.
\end{itemize}

\newpage
\begin{thebibliography}{30}
\bibitem{ref1} P. W. Anderson, Phys. Rev. {\bf 109} (1958) 1492.

\bibitem{ref2} For an overview see e.g. Localisation-1990, edited
by  J. T. Chalker,  IOP
Conf. Proc. No. {\bf 108}  (Institute  of  Physics  and  Physical  Society,
London, 1991).

\bibitem{ref3} D. J. Thouless, Phys. Rep. {\bf 13} (1974) 93;
D. J. Thouless (1979) in  {\em
Ill Condensed Matter}, edited by G. Toulouse and  R. Balian  (North-Holland,
Amsterdam) p.1.

\bibitem{ref4} E. Abrahams, P. W. Anderson, D. C. Licciardello and
T. V. Ramakrishnan,
Phys. Rev. Lett. {\bf 42} (1979) 673.

\bibitem{ref5} P. A. Lee and T. V. Ramakrishnan, Rev. Mod. Phys.
{\bf 57} (1985) 287.

\bibitem{ref6} N. F. Mott and W. D. Twose, Adv. Phys. {\bf 10} (1961) 107.

\bibitem{ref7}  M. Ya. Azbel,  Solid  State  Commun.  {\bf  45} (1983) 527;
Phys. Rev. B{\bf 28} (1983) 4106.

\bibitem{ref8} J. B. Pendry, J. Phys. C {\bf 20} (1987) 733.

\bibitem{ref9} D. H. Dunlap, H. L. Wu  and  P. Phillips,
Phys. Rev. Lett.  {\bf  65} (1990) 88;
P. Phillips, H. L. Wu and D. H. Dunlap, Mod. Phys. Lett. B {\bf 4} (1990)
1249.

\bibitem{ref10} X. C. Xie and  S. Das  Sarma,  Phys. Rev. Lett.
{\bf  60} (1988) 1585;  S. Das
Sarma, He Song and X. C. Xie, Phys. Rev. B {\bf 41} (1990) 5544.

\bibitem{ref11}  Arunava  Chakrabarti, S. N. Karmakar  and   R. K. Moitra,
Phys. Lett. A {\bf 168} (1992) 301.

\bibitem{ref12} J. Q. You, J. R. Yan, T. Xie, X. Zeng and J. X. Zhong,
J. Phys: Condens. Matter {\bf 3} (1991) 7255.

\bibitem{ref13} S. Sil, S. N. Karmakar, R. K. Moitra and  Arunava
Chakrabarti, Phys. Rev. B (In Press).

\bibitem{ref14} J. X. Zhong, T. Xie, J. Q. You and J. R. Yan,
Z. Phys. B: Condensed Matter {\bf 87} (1992) 223.

\bibitem{ref15} D. J. Thouless, J. Phys. C {\bf 5} (1972) 77.

\bibitem{ref16} B. W. Southern, A. A. Kumar, P. D. Loly and
A. M. S. Tremblay, Phys. Rev. B {\bf  27} (1983) 1405.

\bibitem{ref17} M. Kohmoto, B. Sutherland and C. Tang,
Phys. Rev. B {\bf 35} (1987) 1020.
\end{thebibliography}
\end{document}